%% file: RunFiles.tex
\documentclass[useAMS,usenatbib]{mn2e}
\pdfoutput=1
\usepackage{enumitem}
\usepackage{natbib}
\usepackage{graphicx}
\usepackage{amssymb}
\usepackage{lineno}
\usepackage{amsmath}

\begin{document}
\include{Define_NewCommands}

\include{MFile1}

\include{References}
\end{document}

%% file: Define_NewCommands.tex
\newcommand{\MSun}{{M_\odot}}
\newcommand{\LSun}{{L_\odot}}
\newcommand{\Rstar}{{R_\star}}
\newcommand{\calE}{{\cal{E}}}
\newcommand{\calM}{{\cal{M}}}
\newcommand{\calV}{{\cal{V}}}
\newcommand{\calO}{{\cal{O}}}
\newcommand{\calH}{{\cal{H}}}
\newcommand{\calD}{{\cal{D}}}
\newcommand{\calB}{{\cal{B}}}
\newcommand{\calK}{{\cal{K}}}
\newcommand{\labeln}[1]{\label{#1}}
\newcommand{\Lsolar}{L$_{\odot}$}
\newcommand{\farcmin}{\hbox{$.\mkern-4mu^\prime$}}
\newcommand{\farcsec}{\hbox{$.\!\!^{\prime\prime}$}}
\newcommand{\kms}{\rm km\,s^{-1}}
\newcommand{\cc}{\rm cm^{-3}}
\newcommand{\Alfven}{$\rm Alfv\acute{e}n$}
\newcommand{\Vap}{V^\mathrm{P}_\mathrm{A}}
\newcommand{\Vat}{V^\mathrm{T}_\mathrm{A}}
\newcommand{\D}{\partial}
\newcommand{\DD}{\frac}
\newcommand{\TAW}{\tiny{\rm TAW}}
\newcommand{\mm }{\mathrm}
\newcommand{\Bp }{B_\mathrm{p}}
\newcommand{\Bpr }{B_\mathrm{r}}
\newcommand{\Bpz }{B_\mathrm{\theta}}
\newcommand{\Bt }{B_\mathrm{T}}
\newcommand{\Vp }{V_\mathrm{p}}
\newcommand{\Vpr }{V_\mathrm{r}}
\newcommand{\Vpz }{V_\mathrm{\theta}}
\newcommand{\Vt }{V_\mathrm{\varphi}}
\newcommand{\Ti }{T_\mathrm{i}}
\newcommand{\Te }{T_\mathrm{e}}
\newcommand{\rtr }{r_\mathrm{tr}}
\newcommand{\rbl }{r_\mathrm{BL}}
\newcommand{\rtrun }{r_\mathrm{trun}}
\newcommand{\thet }{\theta}
\newcommand{\thetd }{\theta_\mathrm{d}}
\newcommand{\thd }{\theta_d}
\newcommand{\thw }{\theta_W}
\newcommand{\beq}{\begin{equation}}
\newcommand{\eeq}{\end{equation}}
\newcommand{\ben}{\begin{enumerate}}
\newcommand{\een}{\end{enumerate}}
\newcommand{\bit}{\begin{itemize}}
\newcommand{\eit}{\end{itemize}}
\newcommand{\barr}{\begin{array}}
\newcommand{\earr}{\end{array}}
\newcommand{\bc}{\begin{center}}
\newcommand{\ec}{\end{center}}
\newcommand{\DroII}{\overline{\overline{\rm D}}}
\newcommand{\DroI}{{\overline{\rm D}}}
\newcommand{\eps}{\epsilon}
\newcommand{\veps}{\varepsilon}
\newcommand{\vepsdi}{{\cal E}^\mathrm{d}_\mathrm{i}}
\newcommand{\vepsde}{{\cal E}^\mathrm{d}_\mathrm{e}}
\newcommand{\lraS}{\longmapsto}
\newcommand{\lra}{\longrightarrow}
\newcommand{\LRA}{\Longrightarrow}
\newcommand{\Equival}{\Longleftrightarrow}
\newcommand{\DRA}{\Downarrow}
\newcommand{\LLRA}{\Longleftrightarrow}
\newcommand{\diver}{\mbox{\,div}}
\newcommand{\grad}{\mbox{\,grad}}
\newcommand{\cd}{\!\cdot\!}
\newcommand{\Msun}{{\,{\cal M}_{\odot}}}
\newcommand{\Mstar}{{\,{\cal M}_{\star}}}
\newcommand{\Mdot}{{\,\dot{\cal M}}}
\newcommand{\ds}{ds}
\newcommand{\dt}{dt}
\newcommand{\dx}{dx}
\newcommand{\dr}{dr}
\newcommand{\dth}{d\theta}
\newcommand{\dphi}{d\phi}

\newcommand{\pt}{\frac{\partial}{\partial t}}
\newcommand{\pk}{\frac{\partial}{\partial x^k}}
\newcommand{\pj}{\frac{\partial}{\partial x^j}}
\newcommand{\pmu}{\frac{\partial}{\partial x^\mu}}
\newcommand{\pr}{\frac{\partial}{\partial r}}
\newcommand{\pth}{\frac{\partial}{\partial \theta}}
\newcommand{\pR}{\frac{\partial}{\partial R}}
\newcommand{\pZ}{\frac{\partial}{\partial Z}}
\newcommand{\pphi}{\frac{\partial}{\partial \phi}}

\newcommand{\vadve}{v^k-\frac{1}{\alpha}\beta^k}
\newcommand{\vadv}{v_{Adv}^k}
\newcommand{\dv}{\sqrt{-g}}
\newcommand{\fdv}{\frac{1}{\dv}}
\newcommand{\dvr}{{\tilde{\rho}}^2\sin\theta}
\newcommand{\dvt}{{\tilde{\rho}}\sin\theta}
\newcommand{\dvrss}{r^2\sin\theta}
\newcommand{\dvtss}{r\sin\theta}
\newcommand{\dd}{\sqrt{\gamma}}
\newcommand{\ddw}{\tilde{\rho}^2\sin\theta}
\newcommand{\mbh}{M_{BH}}
\newcommand{\dualf}{\!\!\!\!\left.\right.^\ast\!\! F}
\newcommand{\cdt}{\frac{1}{\dv}\pt}
\newcommand{\cdr}{\frac{1}{\dv}\pr}
\newcommand{\cdth}{\frac{1}{\dv}\pth}
\newcommand{\cdk}{\frac{1}{\dv}\pk}
\newcommand{\cdj}{\frac{1}{\dv}\pj}
\newcommand{\rad}{\;r\! a\! d\;}
\newcommand{\half}{\frac{1}{2}}

%% file: MFile1.tex
\title[
Glitches: the exact quantum signatures of pulsars metamorphosis]
{Glitches: the exact quantum signatures of pulsars metamorphosis{}}
{}
\author[Hujeirat,  A.A.]
       {Hujeirat, A.A. \thanks{E-mail:AHujeirat@uni-hd.de} \\
\\
IWR, Universit\"at Heidelberg, 69120 Heidelberg, Germany \\
}
\date{Accepted  ...}

\pagerange{\pageref{firstpage}--\pageref{lastpage}} \pubyear{2002}

\maketitle

\label{firstpage}

\begin{abstract}
 The observed recurrence of glitches in pulsars and neutron stars carry rich information about the evolution of their internal structures.

  In this article I show that  the glitch-events observed in pulsars are exact quantum signatures for their metamorphosis into dark super-baryons (SBs), whose interiors are made of purely incompressible superconducting gluon-quark superfluids.
  Here the quantum nuclear shell model is adopted to describe the permitted energy levels of the SB, which are assumed to be identical to the discrete spinning rates $\Omega_{SB},$ that SBs are allowed to rotate with.
  Accordingly, a glitch-event corresponds to a prompt spin-down of the superconducting SB from one energy level to the next, thereby expelling a certain number of vortices, which in turn  spins-up the ambient medium. The process is provoked mainly by the negative torque of the ambient dissipative nuclear fluid and by a universal scalar field $\phi$ at the background of a supranuclear dense matter. As dictated by the Onsager-Feynman equation, the prompt spin-down must be associated with increase of the dimensions of the embryonic SB  to finally convert the entire pulsar into SB-Objects
  on the scale of Gyrs.

   Based on our calculations, a Vela-like pulsar should display billions of glitches during its lifetime,  before it metamorphoses entirely into a maximally compact SB-object and disappears from our observational windows.
   The present model predicts the mass of SBs and $\Delta \Omega/\Omega$ in young pulsars to be relatively lower than their older counterparts.
  \end{abstract}


\section{Internal structure of pulsars}
   Besides other puzzling activities, the glitch-phenomena observed  in pulsars and young neutron stars are considered to be the key for decoding their internal structures. The
   energies associated with their prompt spin-up are vast and in most of the cases beyond $10^{40}$ erg/event \citep{Haensel2007,Espinoza2011,Eya2014}. However, these objects are extraordinary compact with predicted compactness parameter $\alpha_S~(\doteq R_S/R_\star) \geq 1/2$ and central densities that are beyond  the nuclear density
   $\rho_0 (\doteq  2.7~10^{14}\textrm{ g/cc})$. Here
   $R_S \doteq  2G~M/c^2$ is the Schwarzschild radius, where $G,\,M,\,c$ stand for the gravitational constant, the mass of the object and the speed of light, respectively.\\
   Although the state of matter in this exotic density regime is neither clear nor verifiable, a transition into a quark phase under normal conditions is considered to be  theoretically unlikely \citep{Baym1976,Chapline1977}.  However, in the presence of  a catalyst, such as a universal scalar field   $-\, \phi$ at the background of supranuclear dense matter, specifically in the density regime $\rho \geq 3\,\rho_0,$ sub-nucleon particles may start to frequently interact with the field$-\, \phi,$ thereby gaining energy and steepening
    the curvature of the embedding spacetime, which in turn, compresses the nuclear matter  together and rendering their merger possible \citep{Hujeirat2017}.
    Indeed, in supranuclear dense fluids new communication channels could form and new mesons must then be created  to communicate the residual of the nuclear forces between quarks
    belonging to different baryons. Had the effective energy of mesons became comparable to the bag energy enclosing quarks, then neutrons
   may start merging together to form a super-baryon.  Although the formation of such short-living exotic particles, such pentaquarks, have been experimentally verified at the  Large Hadronic Collider (-LHC) during the years 2009-12, we expect the physical conditions governing
   the supranuclear dense matter at the center of pulsars and neutron stars (-NSs) to lengthen  their long-term lifetimes.

     In fact the rest energy of pentaquarks detected by the LHC was found to lay between 4.38 - 4.45 GeV \citep{LHCb2015}, which is much larger than the sum  of just two individual baryons. Hence a significant part of this energy is stored in the newly formed  communication channels through which  the strong force is communicated.
     For example, viewing the gluon-cloud inside a baryon as a collection of discrete bonds connecting quarks, then the energy  of a super-baryon consisting of hexaquarks would have 15 bonds and therefore $15  \times (0.938/3)~ GeV\,\approx \,4.6\,GeV,$ which is  relatively close to the revealed values.

\begin{figure}
\centering {\hspace*{0.75cm}
\includegraphics*[angle=-0, width=4.15cm]{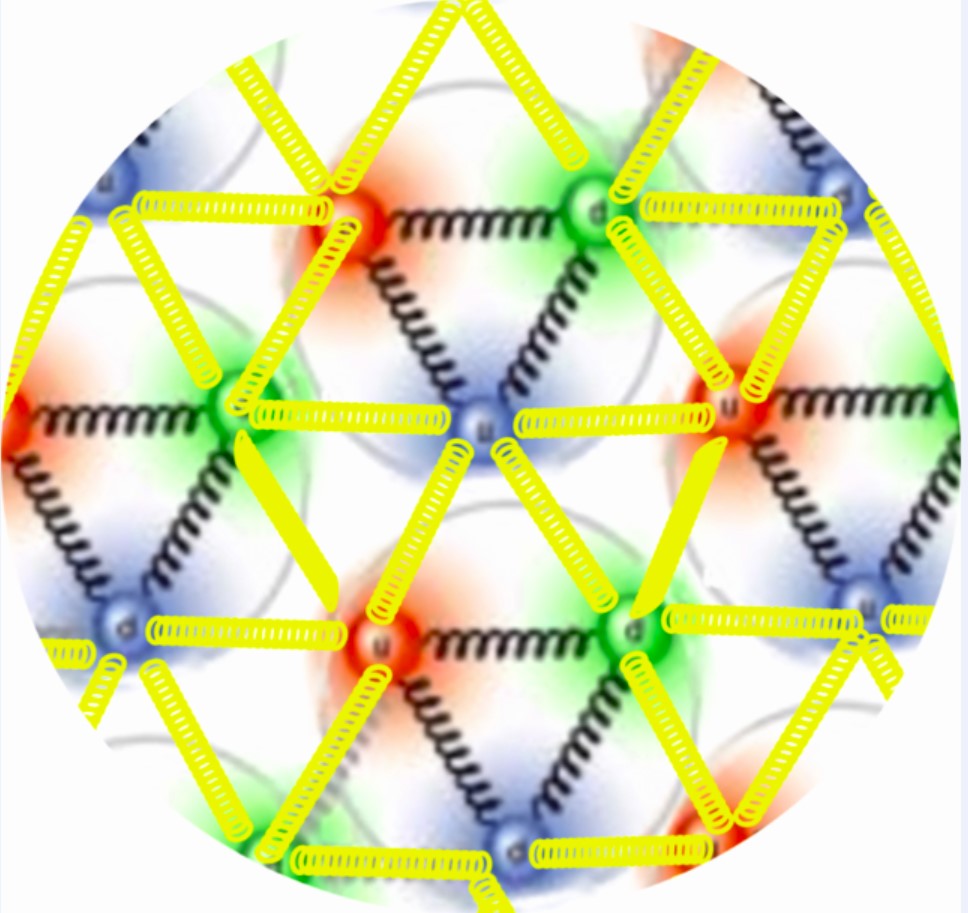}
}
\caption{\small  A two-dimensional illustration of strongly compressed baryons at the center of pulsars and neutron stars. Each baryon here consists of three quarks (red, blue and green), that are coupled together via the strong nuclear force, i.e. the gluon-field  (black-colored bonds). When baryons are sufficiently compressed together, then numerous new communication channels between quarks belonging to different baryons would form (yellow-colored bonds) rendering merger of baryons possible and forming a super-baryon.
}  \label{XXX}
\end{figure}

      Recalling that the bulk mass of a baryon mass originates mainly from the gluon-cloud embedding the quarks, then
      the number of gluon-channels connecting $ 10^{57}$ quarks inside a stellar-size
      super-baryon would increase dramatically (Fig. 1) and would ensue a self-collapse
      of the SB into a black hole. However, as neither BHs nor NSs ever observed in the mass-range
      $[2\MSun \leq \calM \leq 5 \MSun]$ and as formation of stellar-size gluon-quark objects cannot be excluded, then the gluon contribution to mass cannot increase super-linearly  with
      increasing the number of quarks involved. Hence a stable pulsar of initial mass  $\calM=\calM_{NS}$ and $\alpha_s=1/2$ would at most double its mass in order to escape its collapse
      into a BH.
  At supranuclear densities, almost all EOSs tend to converge to the limiting case:
  $P\rightarrow \calE = a_0 \check{n}^2,$ where $\check{n}$ is the neutron number density \citep{Camenzind2007,Glendenning2007}. Due to causality reasons, the chemical potential $\mu$ cannot grow indefinitely and it must be upper-bounded by $\mu_0,$ beyond which the fluid
  acquires a constant $\mu$ and therefore a constant $"\check{n}"$; practically becoming a purely incompressible nuclear fluid \citep{Hujeirat2009,Hujeirat2017}. This
  implies that neutrons at the center of pulsars would start merging together to form a super-baryon, whose interior
   is made of a continuum with a uniform energy density. The scalar field here enters the  process as a pseudo-catalyst:
 \beq
  \calE = a_0 \check{n}^2  \xrightarrow{\textrm{dark energy}}   \calE = a_{qsf} \times \check{n},
 \eeq
 where $a,~a_{qsf}$ are constant coefficients.  \\
 In a pervious article \citep{Hujeirat2017}, it was shown that the onset of phase transition
 into the quark-phase occurs at $\rho= \rho_{cr}=3\times\rho_0,$ at which the entropy vanishes and the Gibbs function attains a global zero-minimum. \\
 In such fluids the  gradient of the local pressure vanishes and replaced by the gradient of a
  global negative pressure generated by the scalar field $-\phi$ \citep{Hujeirat2017}. The injected dark energy here was found to be equal to the amount
  required for deconfining the enclosed quarks, but also to the energy needed
  to increase $\alpha_S$ from $1/2$ to approximately one: the critical value at which the object sinks deeply into the embedding strongly curved spacetime and
  becomes invisible.\\

 \noindent In Figures (2) and (3) the basic ingredients  of the present model are visualized. Accordingly,
 when a pulsar is born, the very central density surpasses the nuclear density $\rho_0$ and settles around the critical density $\rho_{cr} = 3\times\rho_0,$ where neutrons, under the
  effect of the enhanced gravitational field, start merging together to form an SB at the center of the pulsar (the central sphere in top Figure 2). Here the matter
  changes its state according to Eq. (1) to become an incompressible superfluid with constant chemical potential.
   With increasing the dimensions of the SB and therefore the number of the enclosed quarks, the density of the embedding gluon-cloud becomes larger\footnote{The gluon cloud can viewed as a collection of flux-tubes connecting quarks, whose number must increases super-linearly relative to the number of quarks.} and yields an enhancement of the effective mass of the SB. In the present study, increasing of the number of flux-tubes
    between quarks is equivalent to injection of dark energy, which accounted for through modifying the EOS in the following manner:  $\calE_{tot} = \calE_{b} + \calE_{\phi},$ where $\phi$ corresponds to the scalar field $\phi.$\\
  Similar to laminar flows over solid static bodies, a thin boundary layer (-BL) between the dissipative ambient medium and the rigid-body rotating SB is predicated to form. The dynamics of the fluid inside the viscous BL, which initially was rotating differentially (due to shear viscosity), would start to be increasingly influenced by the rest of the strong nuclear force in combination with compression-enhancement due to the injection of dark energy. These forces would quickly start to dominate over viscous stresses to subsequently provoke its transition into the superfluid phase. The matter inside the BL
  would lock to the SB  and starts rotating coherently with it.\\
  Our model is based on the assumption that the SB behaves as a single quantum entity,
  whose permitted rotational frequencies form a sequence of discrete values $\{\Omega^n_{SB}\}.$\\
  Once the rotational frequency of the ambient medium has decreased  and hit
  a suitable element of $\{\Omega^n_{SB}\},$  then the SB undergos a sudden spin-down and
  starts rotating with this lower  value, thereby ejecting a  certain number of vortices.
  In turn, these vortices are absorbed then by the ambient dissipative medium and provoke its prompt spin up, which mimics a single glitch event out of many that have been observed to associate the long-term evolution of pulsars and young neutron stars.
  Each time the SB undergoes a spin-down, its size and inertia must increase respectively to finally metamorphose the entire pulsar into a stellar-size SB. At that moment, the object should have
  attained its  maximum compactness and would turn invisible completely.
\begin{figure}
\centering {\hspace*{0.75cm}
\includegraphics*[angle=-0, width=7.15cm]{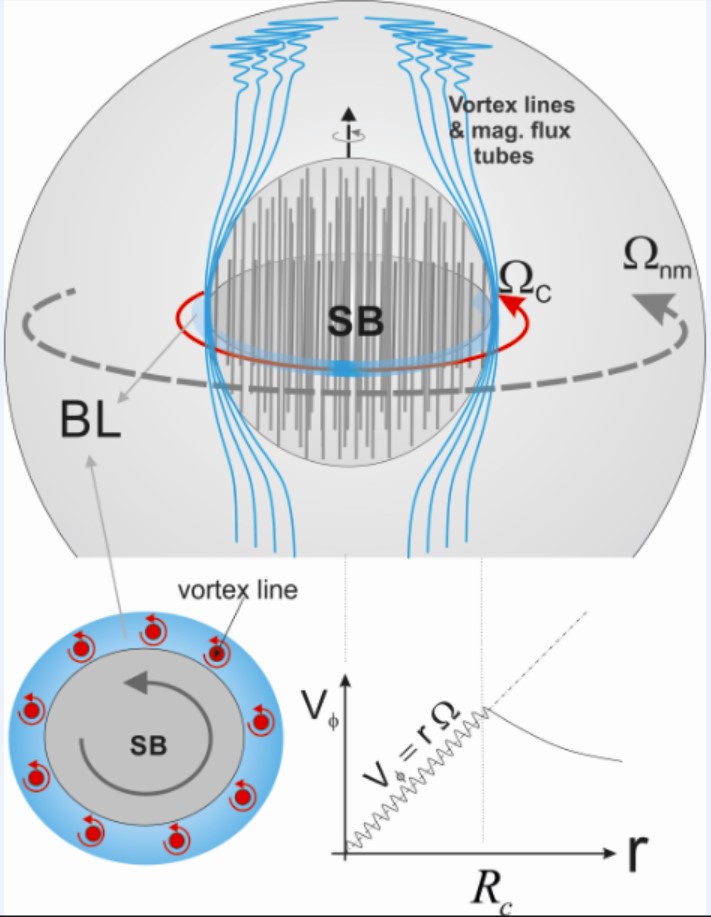}
}
\caption{\small A schematic description of a pulsar enclosing an embryonic
super-baryon (SB); whose interior is made of superconducting gluon-quark superfluid  surrounded by a geometrically thin boundary layer (BL). According to the
present scenario, the ambient medium is made of a dissipative nuclear matter
that rotates differentially i.e., $d\Omega_{nm}/dr <0$. The interior of the SB
is governed by numerous mini-vortices whose vortex-density is uniform and increases linearly
with radius, thereby giving rise to  rigid body rotation with $\Omega_{SB}.$ The BL here is made of neutron-superfluid and threaded by a time-dependent number of vortex lines that rotate coherently with the SB.
}  \label{XXX}
\end{figure}

  \section{Signal transmission from evolving super-baryons in pulsars?}
  Assume the Vela pulsar to have been born with the initial angular velocity $\Omega_0 = 1540/$sec and magnetic field $B_0=10^{13}~$ Gauss. If the entire interior of the pulsar were made of a superfluid and its total rotational energy were
  stored in a network of vortex lines,  then we may use Onsager-Feynman equation,
  $\oint \textbf{v}\cdot \textbf{d} l = \DD{2\pi \hbar}{ m}N_V $ to compute  the total number of vortex line threading the entire object. Here  $N_V,~\textbf{v},\textbf{d}l,~ \hbar, m$ denote the total number of vortex lines, velocity field, vector of line-element, the reduced Planck constant and  the mass of the superfluid particle pair, respectively. Assuming the number density of vortices across the entire object to be constant, then the enclosed number
  of vortex lines inside a sphere of radius "r" would be $N_V(r) = const. \times r^2,$ which mimics rigid body rotation with $\Omega = const.$\\
  However, this picture oversimplifies the problem and leaves numerous questions open, in particular those related  to the origin, dynamics and timing of glitches as well as to the physics and EOSs of  supranuclear dense superfluids.\\

  Similar to normal stars, the cores of pulsars are expected to initially consist of normal dissipative fluids. Under the effect of gravity and exotic fields, neutrons at the very central region would start merging together to form an embryonic SB. The inertia of the SB relative to the entire star behaves like $I_{SB}/I_\star \sim (R_{SB}/R_\star)^5.$ Thus, unless the SB is of a considerable macroscopic size, its dynamics would be completely insignificant. \\
  Inside the rotating SB, the number density of the mini-vortices is constant and therefore the SB can be safely considered as a rigid body rotator (see Fig. 2).

  Consider a terrestrial ${}^4$Helium-superfluid  in a container.  It was  experimentally
  verified that the superfluid would not change its rotational frequency, unless its rotational frequency has hit a term, which the superfluid is allowed to spin with in
    accord with the laws of quantum mechanics \citep{Yarmchuk1979}. Analogously, the SB-pulsar system
  can be viewed as follows:
  \noindent \hfill
\beq
\barr{lll}
  {}^4\textrm{Helium superfluid} & \leftrightarrow & \textrm{superfluid inside the SB}\\
   \textrm{ Solid container}       & \leftrightarrow & \textrm{Ambient dissipative medium}\\
                          &  & \textrm{surrounding the SB}.
\earr
\eeq

Here the emission of magnetic dipole radiation enforces the ambient medium to spin-down continuously. While spinning down, the ambient medium certainly would hit an element of the $\Omega_c-$sequence, whose elements are the values that the SB is allowed to rotate with.
In order to lower its spin, the SB must expels instantly a certain number of vortices.
However, the ambient dissipative medium would then absorb the expelled vortices and spin-up
promptly as illustrated in Fig. (3) and calculated in Fig (5). \\

\begin{figure}
\centering {\hspace*{0.75cm}
\includegraphics*[angle=-0, width=7.0175cm]{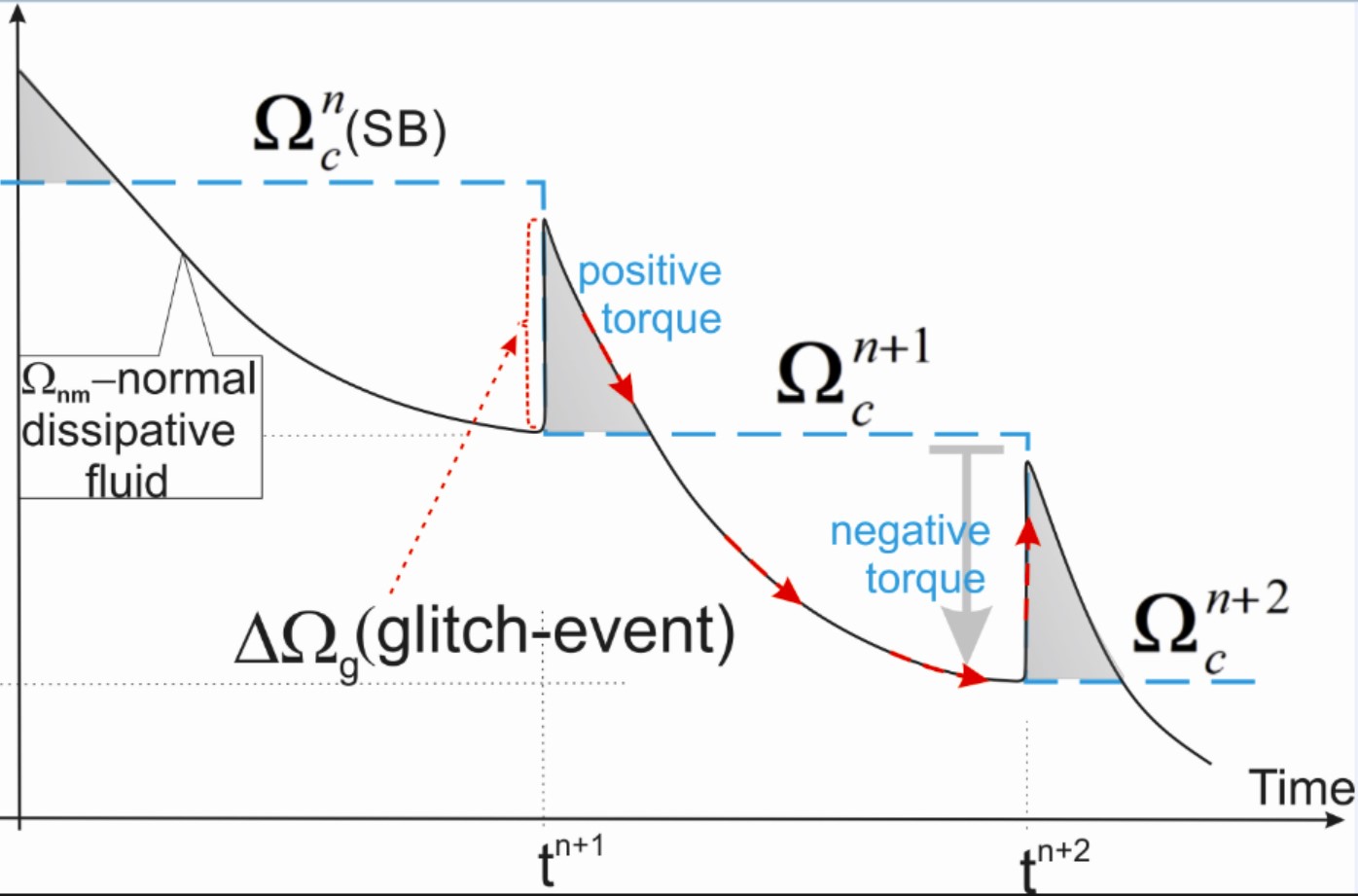}\\
\includegraphics*[angle=-0, width=7.0175cm]{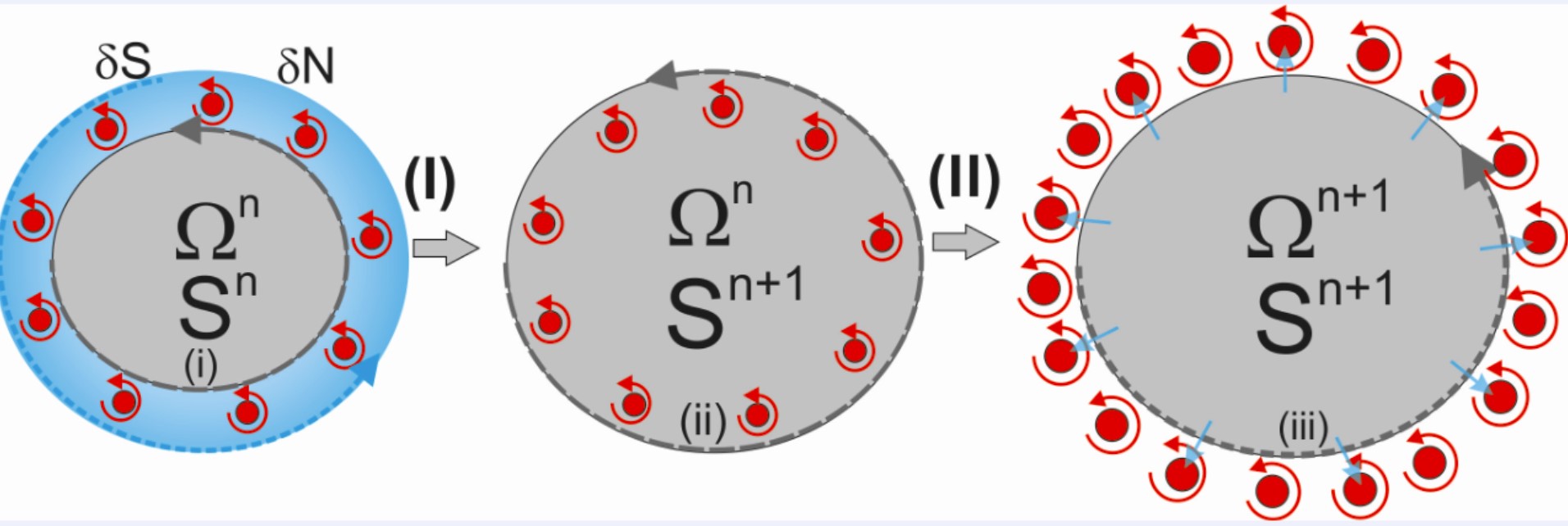}\\
\includegraphics*[angle=-0, width=7.0175cm]{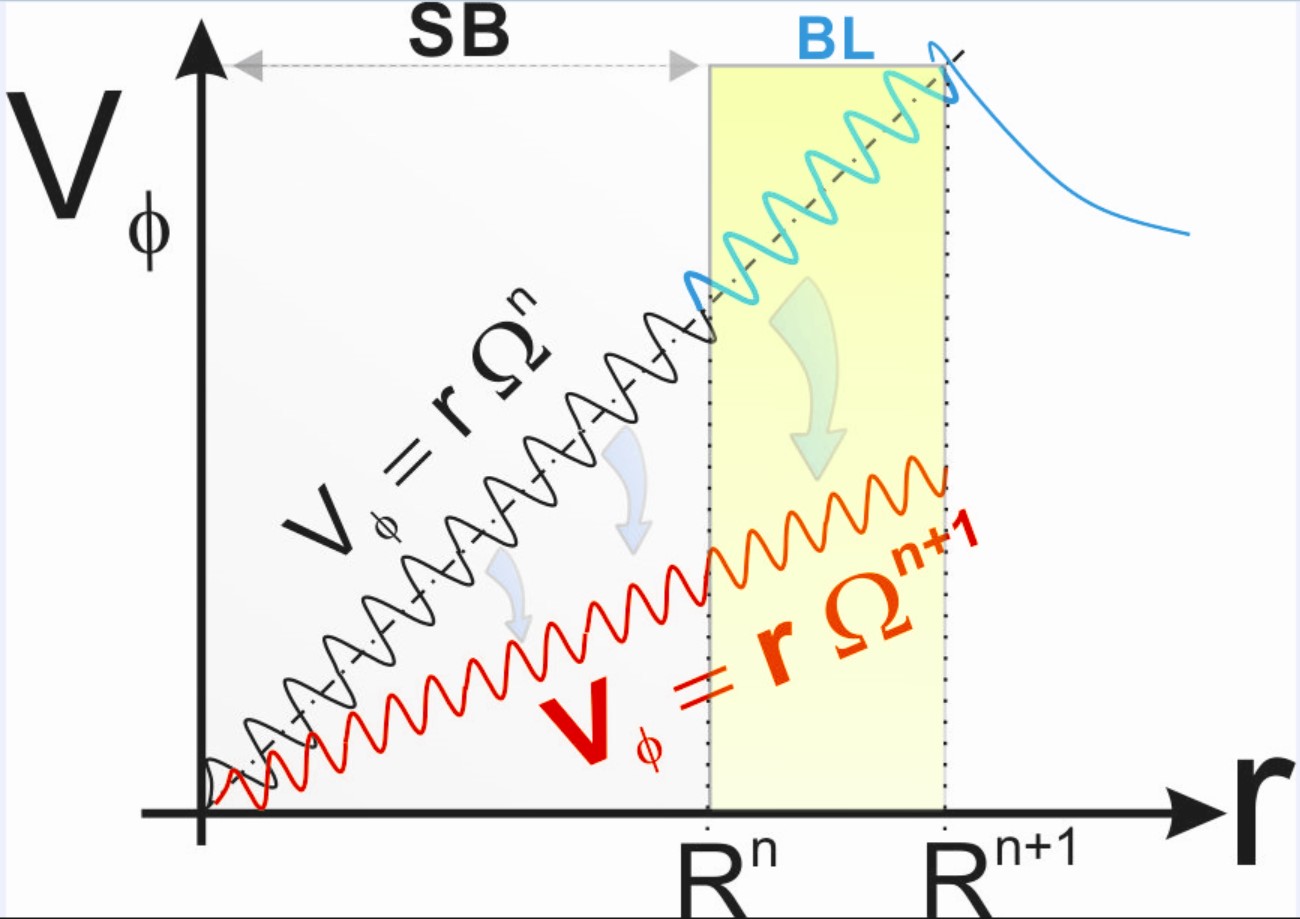}
}
\caption{\small A schematic description of a glitch event of a pulsar enclosing an embryonic super-baryon versus time. Within the period
 $\delta \tau^n_c = [t^{n+2} - t^{n+1}],$ the SB rotates rigidly with a  constant
  $\Omega_{SB},$ hence the corresponding angular velocity is
  $V_\phi= r \Omega^n_{SB}.$ During $\delta \tau^n_c$ the BL starts to form and the residual nuclear force starts to dominate over viscous forces, thereby converting the matter into the superfluid phase and enforcing its vortices  to rotate rigidly and coherently with the SB (bottom panel/blue colored line). On the other hand the torque exerted  by the dipole magnetic field enforces the rotational frequency of the ambient dissipative medium to spin-down from $\Omega^n_{c}$  to  $\Omega^{n+2}_c$ continuously.  Within $\delta \tau^n_c,$
the SB-BL system finds itself rotating increasingly faster than its ambient dissipative counterpart.
Once  the angular frequency of the surrounding medium, $\Omega_{nm},$  has fallen below a certain critical value
$\Omega_{crit},$
the rotational frequency of the SB-BL system jumps promptly
into the next lower and permitted value $\Omega^{n+2}_c,$ where now it rotates rigidly
with the angular velocity $V_\phi= r \Omega^{n+1}_{SB}$ (the red-profile in the bottom panel). The $\Omega-$jump here is
associated with a prompt increase of the mass, energy and inertia of the SB.
}\label{XXX}
\end{figure}

  To follow the time-evolution of the interactions between the BL, SB and the ambient medium,
  the Onsager-Feynman equation (OFE) is employed. In the finite and flat-spacetime the time-derivative  of the OFE reads:
 \[
\DD{(S\Omega)^{n+1} - (S\Omega)^{n}}{\Delta t}  =
  \DD{h}{ 2m} (\DD{N^{n+1}-N^{n}}{\Delta t}),
\]
where $\Delta t (\doteq t^{n+1} - t^n)$ is a time-interval between two successive glitch events.  Applying the equation to the step function of $\Omega_c$  (see Fig. 3), then the equation gets the form:

  \beq
  (S\Omega)^{n+1} - (S\Omega)^{n} = [S~\Omega] = (h/2m) ~\delta N_V.
  \eeq
  This equation states that the matter in the concerned region is allowed to
   change its state  from $(S\Omega)^{n}$ into $(S\Omega)^{n+1},$  if a certain number of vortices, $\delta N,$ is extracted out of the system.\\
  Consider now the matter inside the BL (see Fig. 2 and 3). Shortly after ejection of
  vortices, this matter must have been rotating differentially, but then starts to be increasingly affected by the residual of the strong nuclear force of the SB, which finally
    becomes dominant over the viscous forces. The force here locks the matter to the SB and enforce its transition from normal viscous neutron fluid state into a superfluid phase, inside
    which vortices develop and rotate coherently with those inside the SB (see Figs. 1, 2 and 3).\\
  Right at the $\Omega-$ discontinuity (e.g.  $t=t^{n+1}$ in Fig. 3), two simultaneous actions are expected  to occur:

  \begin{itemize}
    \item   As the matter both in the BL and in the SB rotates rigidly with the same $\Omega^n,$ then the SB-BL system  is rotationally behaving as a single quantum entity.
        In this case the corresponding OFE reads:
     \[
           S^{n+1}\Omega^{n} = (h/2m)N^*_V,
    \]
          where $N^* = (S^{n+1}/S^n)N^{n} (> N^{n})$ is the total number of vortices enclosed in the SB-BL system (see sphere (ii) in Fig. 3). In writing the expression for $N^*,$ the rigid body rotation of the matter in the BL is taken into account and that the number densities of vortices in both regions are equal.\\

    \item The matter inside the modified SB, namely inside $S^{n+1},$
          starts to interact directly with the surrounding viscous matter, which rotates with the lower frequency $\Omega^{n+1};$ an element of $\Omega-$sequence  which the SB is allowed to quantum-mechanically rotate with. Hence the SB spin-down promptly from $\Omega^{n}$ into $\Omega^{n+1},$ thereby ejecting
          a certain number of vortices that amounts to: \\
          \[
             \delta N_V = N_V^* - N^{n+1} =[N_v]= (2m/h) S^{n+1}(\Omega^{n}-\Omega^{n+1}).
          \]
           The  loss of rotational energy here is assumed to be the mechanism that triggers the phase transition of the superfluid neutrons inside the LB into an incompressible gluon-quark superfluid state,
           where the EOS changes from $\calE = a_1 \check{n}^2$ into  $\calE = a_2 \check{n}$ and where $a_1, a_2$ are constants and $\check{n}$ is the number density of matter.
           The expulsion of this energy would certainly affect the dynamical stability of the SB and in principle could provoke a dynamical collapse of the SB.
           This collapse may still be escaped  if the SB enlarges its  dimensions in such a manner that its total rotational energy is conserved, i.e.:
            \[
             E^n_{rot} = \DD{1}{2}I^n_{SB}(\Omega^n)^2 = \DD{1}{2} I^{n+1}_{SB}(\Omega^{n+1})^2= E^{n+1}_{rot}
           \]
            Recalling that the matter inside the SB has the uniform constant  density $\rho=\rho_{cr},$   then the following relation can be obtained:
          \beq
                           S^{n+1} = S^{n}(\DD{\Omega^{n}}{\Omega^{n+1}})^{\Gamma},
          \eeq
           where $\Gamma$ is a constant.
             Depending on the density of dark energy,  $\Gamma$ may
             vary from $\Gamma =4/5$ in the case of pure baryonic matter up to
              $\Gamma \approx 1$
             in dark energy dominated case. \\
             On the other hand, as $\Omega^{n}/\Omega^{n+1} = 1 + \epsilon,$ where $\epsilon\ll 1,$   I use, for simplicity, the limiting case: $S^{n+1} = S^{n}(\Omega^{n}/\Omega^{n+1})$ to predict the
             size of the SB shortly after each glitch-event.
  \end{itemize}

   To conclude: when applying the OFE to $\Omega-$discontinuity at $t=t^{n+1},$ then the
   equation may be split into the following two parts:
    \beq
  [S\Omega] = (h/2m) \delta N_V   \Rightarrow
         \left\{
       \begin{array}{ll}
              i. & [S\Omega]=0     \\
              ii. & S[\Omega] = (h/2m)~ [N_V].
             \end{array}
  \right.
  \eeq
  Hence,  each time the pulsar undergo a glitch event,  the dimensions of the SB
  should increase and similarly its gravitational mass and inertia. The ambient medium
  however  would  experience an opposite development:  it would become increasingly smaller, dynamically insignificant to finally vanish completely.\\

  Since the discrete energy levels of an SB embedded in a strongly curved spacetime
  are completely unknown, I use  pulsar's observation to guide the search for a reliable  $\Omega-$sequence,
  whose elements are likely to be the values that SBs are allowed to rotate with. \\
  In the present study, instead of $\{\Omega_c^n\},$ I use for simplicity the sequence $\{\alpha_c^n\},$ whose elements are defined as follows:
   \[
    \alpha^n_c= {\Omega^{n}_c}/{\Omega^{n+1}_c} = 1 + (\Delta\Omega/\Omega)^{n},
   \]
   where the terms $\{(\Delta\Omega/\Omega)^n\}^\infty_{n=1}$ are observables  that can be used to constraint the $\alpha_c^n-$sequence. Moreover, in searching for the best suited sequence, the  two limiting cases of the cross section of the SB should be taken into account:
         \beq
            \barr{lllll}
           i.&~~~~  S_c(t=0) &= & S^0   \\
          ii.&~~~~  S_c(t\rightarrow\infty)  &= & S^\infty= S^\star,  \\
      \earr
            \eeq
\noindent where $S^0, S^\infty$ denotes the initial and final cross-sections of the
dynamically evolving SB, respectively.

  Noting that ${\Omega^{n}_c}/{\Omega^{n+1}_c} $ is a slowly varying function with time (e.g. on the scales  of years), the following two relation can be obtained:
    \beq
    \barr{lll}
        S^{n}_c =  & ( \alpha^0_c \times  \alpha^1_c   \times  \alpha^2_c  ... \alpha^n_c ) ~S^0   = (\displaystyle\prod_{i=0}^{n}\alpha^i_c) ~S^0_c, &\\
       & \textrm{and}  &\\
      & &\\
        \Omega^{n}_c = &  ~\Omega^0 /( \alpha^0_c \times  \alpha^1_c   \times  \alpha^2_c  ... \alpha^n_c ) = ~\Omega^0 /(\displaystyle\prod_{i=0}^{n}\alpha^i_c). &
       \earr
  \eeq
   Using theoretical considerations in combination with glitch observation of pulsars, in particular those of  the well-studied Crab and Vela pulsars, a list of constraints may be constructed that can be used to determine the elements of the $\alpha_c^n-$sequence:
     \beq
  \barr{ll}
    1. & \textrm{The observed spin-down of pulsars implies that } ~\alpha^n_c > 1.\\
    2. & \textrm{The Vela pulsar is roughly 10 times older than the}\\
    & \textrm{Crab pulsar}\\
    3. & \alpha^n_c \rightarrow 1 + 2\times 10^{-8}  \textrm{ as }  n \rightarrow \tilde{N}^{Crab} \textrm{ and }\\
       & \alpha^n_c \rightarrow 1 + 2\times 10^{-6}  \textrm{ as }  n \rightarrow \tilde{N}^{Vela},
       \textrm{ where $\tilde{N}^{[*]}$}\\
       &  \textrm{denotes the total number of glitches that the pulsar has} \\
       & \textrm{experienced  so far.} \\
       & \textrm{Accordingly, $\tilde{N}^{Crab}$ must be much smaller than $\tilde{N}^{Vela}$.} \\
    4. & \alpha^n_c \rightarrow 1  \textrm{ as }  n \rightarrow 0: \textrm{This is a consequence of the basic} \\
       &  \textrm{assumption that SBs are initially of nucleon-sizes}\\
       &  \textrm{ and rotating uniformly with the ambient media.}\\
    5. & \displaystyle\prod_{i=0}^{\tilde{N}(Crab)}\alpha^i_c = \Omega^{0}_c/\Omega^{\tilde{N}}_c  =  7.7~~ and
    \displaystyle\prod_{i=0}^{\tilde{N}(Vela)}\alpha^i_c = \Omega^{0}_c/\Omega^{\tilde{N}}_c  =  22\\
    6. & \displaystyle\prod_{i=0}^{n\gg \tilde{N}}\alpha^i_c  = S^{\infty}_c/S^0_c \\
    7. & \textrm{When applied to the Crab and the Vela pulsars, the}  \\
       & \textrm{lapse of time between two successive glitch-events at} \\
       & \textrm{the present time should be of order two years, i.e. }\\
       & \delta t^n = t^{n+1} - t^n \approx 2\,\textrm{years}.\\
  \earr
    \eeq
       Constraint number 1 (henceforth C1) means that pulsars are observed to spin down, i,e, $d\Omega/d\,t <0$ and therefore $\Omega(t-\epsilon)>\Omega(t+\epsilon),$ which is equivalent to $\Omega^n>\Omega^{n+1}.$ C2 means that the element of the sequence should yield the ages of two pulsars correctly. Ages of objects are obtained by summing over the intervals between successive glitches up to the present time, e.g.:
         \[
            \tau^{N(Crab)}_{tot} = \sum^{N(Crab)}_{n=1} \delta \tau^n_c \approx 10^3\,yrs,\]
         and
         \[
            \tau^{N(Vela)}_{tot} = \sum^{N(Vela)}_{n=1} \delta \tau^n_c \approx 10^4\,yrs,
         \]
         where $\delta\tau^n_c$ is calculated using Eq. (11). C3 means that the present observations of the Crab and Vela pulsars  reveal $\Delta \Omega/\Omega$ equal to  $2\times 10^{-8}$ for the Crab and $2\times 10^{-6}$ for the Vela. Hence
         the elements of $\alpha_c-$sequence and the related parameters should be carefully constructed, so that the time intervals between
         two successive glitches at the present time should be  equal to two years approximately and that their metamorphosis processes  should be completed within the first
         hundred million years after their birth.
       C4 implies that   at time$\,=0,$ i.e. when the pulsar was born, the size of the SB was comparable to those of nucleons in atomic nuclei.
       C5 means that the multiplications of the first $N(Crab)$ elements of the
       sequence \{$\alpha^n_c$\} should yield 7.7 in the Crab case, whereas
       the multiplications  of the first
       $N(Vela)$ elements should yield  22 in the Vela case.
       Here I rely on the works by \cite{Hansel1999,Eya2014} to fix the initial frequencies of the pulsar  and on observations to fix the present rotational frequencies of both pulsars. C8 implies that after a very large number of glitches, the final size of the SB
       should be equal to that of the pulsar at its birth. The underlying assumption is here that the total volume, i.e. the volume of the SB plus the volume of the ambient medium, is time-independent and remains   constant during the whole metamorphosis process (see Fig. 4).

\begin{figure}
\centering {\hspace*{0cm}
\includegraphics*[angle=-0, width=8.5cm]{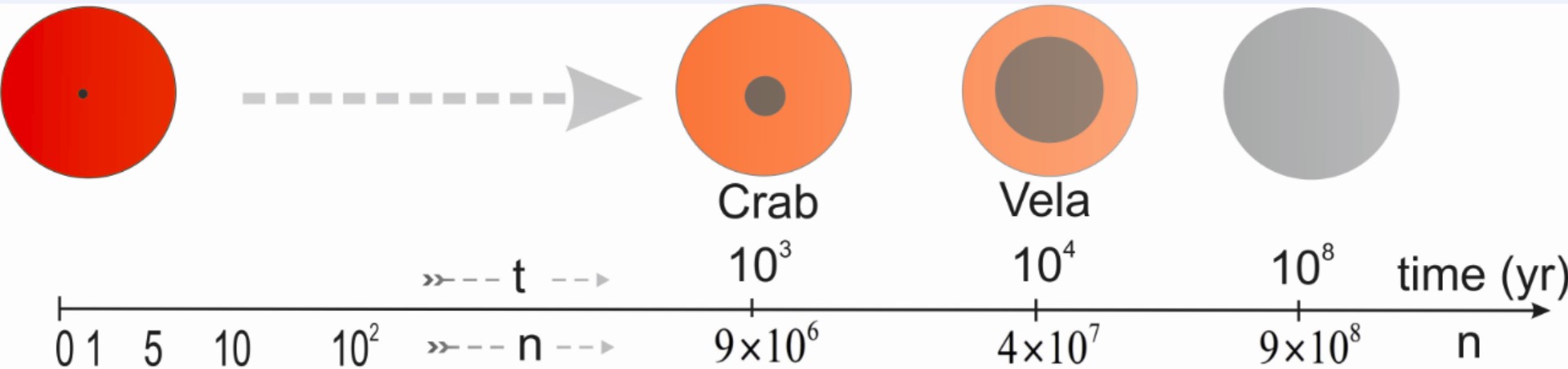}
}
\caption{\small The correlation of the element-number $n$ of $\alpha_c-$sequence, the real time (t)
and the sizes of both the SB (black color) and the surrounding shell of dissipative matter
 (red color) are shown.
}\label{XXX}
\end{figure}

    \noindent Note that since  $\alpha^{n}_c/\alpha^{n-1}_c > 1,$ then the corresponding sequence must not converge and therefore the above-mentioned 6th-constraint
    cannot be fulfill; specifically that the final cross section of the SB should be upper-bounded by that of the pulsar, i.e.
    \[
    S^n \xrightarrow[n \rightarrow \infty]{}  S^\star   ~~~~\small{\bigcap}~~~~~ S^\infty \leq S^\star.
    \]
    As it will be shown in the next section, this problem can be circumvented by allowing the dipole magnetic field to asymptotically decay to zero at the end of the pulsar's lifetime. In that case,  the time between two successive glitch-events would stretch to  infinity (see e.g. Eq. 10 and Fig. 7).\\
  On the other hand, constructing a convergent sequence to describe the
 evolution of glitches in pulsars is not only physically inconsistent, but it is also contradictory to the observations  of  the Crab and Vela pulsars.
One possible improvised sequence that may fulfill most of the above-mentioned constraints
may read as follows:
\beq
    \alpha^n_c  = (1 + \epsilon_1 n^{\epsilon_2})^{\epsilon_3 n^{1+\epsilon_4}}, \textrm{   for } n:~ 0 \rightarrow ~\infty,
\eeq
where $\epsilon_1=10^{-12},~\epsilon_2=5 \times 10^{-2}, \epsilon_3=1+  10^{-10}$ and $\epsilon_4=2 \times 10^{-2}.$\\
Indeed, $\epsilon_i,\,{i=1\rightarrow 4}$ have been  carefully chosen and optimized so to agree with glitch-observation
in pulsars, and in particular to those of the Crab and Vela pulsars.\\
Using the elements of $\{\alpha^n_c\},$ a sequence for the cross sections of the SB can be constructed, $\{S^n_c\},$ from which the sequence $\{R^n_c\}$ is then calculated to determines the radii of the permitted energy levels.\\
\noindent Assuming glitch-events in pulsars to evolve according to a well-defined "magic-"sequence, then the here-presented $\alpha-$sequence may be  considered as one possible
approximation, though additional modification may still be required.

\section{Governing equations \& solution procedure}

   The set of equations to be solved here reads as follows:

\noindent \hfill
\beq
\barr{llll}
  i. & {d}(I \Omega)_{nm}/{dt} & = &
                -\beta B^2 \Omega_{nm}^3 + \delta(\Omega_{nm}-\Omega_c) \Delta \dot{I}_{SB} \Omega_{SB}\\
  ii. & \textrm{$[S \Omega$]} & = &0  ~\textrm{\&}~ S[\Omega]= (h/2m)[N_V] \\
 iii. &  M_{SB} & = & 2\pi \int^{r_c}_0 (\rho_{cr} + \calE_\phi) r^2 dr \\
 iv. &  M_{nm} & = & M_\star - M^b_{SB} \\
 v. &  M_{SB}&  \rightarrow & M_{Sch} \textrm{     as      } n\rightarrow \infty,
  \earr
  \eeq
  \noindent $I_{nm},~ \Omega_{nm},~ B$ in Eq. (i/10) denote the inertia,
  angular frequency and the magnetic field of the normal dissipative medium surrounding the SB.  This equation describes the conservation  of the total angular momentum of the  ambient medium subject to the two forces:
  \bit
  \item The first term on the RHS of Eq. (i/10) stands for the magnetic torque, which acts to brake the rotation of the ambient medium through emission  of magnetic dipole radiation
  \item  The second term on the RHS of Eq. (i/10) represents the rotational energy gained by absorbing the vortices ejected by the SB during a glitch event.
    \eit
    The delta function on the RHS of Eq. (i/10) is defined as follows:
   \[
   \delta(\Omega-\Omega_c) =\begin{cases} 1 \textrm{~~~~  if  } \Omega=\Omega_c \\
                                      0 \textrm{~~~~  otherwise}.
    \end{cases}
   \]
   Equation (ii/10) describes the prompt reaction of the SB as clarified in the previous section.
   Here $[N_V] = \delta N_V \propto \Delta \Omega_{SB} $ represents the prompt loss of rotational energy from the SB-BL system during a glitch-event.
   Once the cross-section of the SB has increased from $S^n$ to $S^{n+1}$, its total effective mass is calculated from Eq. (iii/10), where I use
   $~\calE_\phi \doteq \alpha_0 r^2 + \beta_0$ to calculate the density of the dark energy.
   Here $\beta_0$ is the canonical energy scale characterizing the effective coupling of quarks
   \citep[see][and the references therein]{Bethke2007}.
   Eq. (iv/10) describes mass-reduction
    of the ambient medium  as a consequence of the mass-growth of the SB, where
    it converges to the corresponding Schwarzschild mass as the number of glitches becomes
    very large (Eq. v/10).\\

  The algorithmic structure of the numerical solution procedure runs as follows:\\
  \noindent
  \begin{enumerate}[leftmargin=*,labelindent=16pt,label=\bfseries \arabic*.]
    \item Construct the sequence $\{\alpha^n_c\}, $
      which fulfills the list of conditions listed in Eq. (8).
    \item Calculate the elements of the sequences  that corresponds to $\{\Omega^n_{SB}\}, $  $\{S^n_{SB}\}, $
       $\{R^n_{SB}\} $ and  $\{I^n_{SB}\} $
    \item Calculate the elements of the sequence that correspond to the total mass of the
    SB, i.e.  $\{M^n_{SB}\}_{tot}$, including dark energy as well as its total inertia $\{I^n_{SB}\}_{tot}.$
    \item Calculate the elements of the sequences that corresponds to the mass $\{M^n_{nm}\}$ and inertia  $\{I^n_{nm}\}$ of the ambient dissipative medium.
    \item Calculate the elements of the sequence $\{\delta \tau^n_c\} $ using a physically reasonable formula that relates the magnetic field decay to the inertia $\{I^n_{am}\},$ taking into account that both the magnetic field $B$ and $I_{nm}$ should decay asymptotically go to zero as $t\rightarrow \infty$ (see Eq. 11).
    \item Compute the continuous decay of $\Omega_{nm}$ of the ambient medium as dictated by
    the emission of magnetic dipole radiation ( Eq. i/10).
\end{enumerate}
\noindent In solving the set of Equations (10), $1540 /s$ is used as an  initial value for $\Omega_c(t=0),$ which is
     in the range proposed by \citep{Cook1994,Hansel1999}. While the final  cross section of the object is approximately $S^{\infty}_c = \calO( 10^{12})~cm^2,$ the initial cross section $S^0$ of the SB may  significantly differ from one pulsar to the other, depending on their initial masses and magnetic fields. As I will show later, I anticipate
        $S^0_{SB}(Crab) \ll S^0_{SB}(Vela).$ \\

\noindent
  Note that within the time interval between two successive glitch events: $[t^{n},t^{n+1}],$ both  the inertia $I_{SB}$ and angular velocity $\Omega_{SB}$  are constant, whilst $\Omega_{nm}$ of the ambient medium decreases continuously as dictated by the loss of magnetic dipole radiation (see $\Omega_{nm}$ in Fig. 3). Hence the time interval between two successive glitch-events can be calculated  from the first equality in Eq. (i/10) as follows:
  \beq
  \delta \tau^n_c = \DD{1}{2\beta} \DD{I}{B^2} \{
              (\DD{1}{\Omega^c_{n+1}})^2 - (\DD{1}{\Omega^c_{n-1}})^2    \}
                \doteq  \DD{1}{2\beta} \DD{I}{B^2} \bar{\bar{\Delta}}.
  \eeq
  Based on test calculations, no reasonable values for $\delta \tau^n_c ,$ that
  could agree with observations were found, unless ${I_{nm}}/{B^2}$ are set
   to decrease with time.\\
Actually the numerical solution procedure is intrinsically sequential. Once the elements of the sequence $\{\alpha^n_c\},$ for $n=0\rightarrow \infty$ are fixed, the determination of the other sequences are straightforward. Since the rotational frequency and inertia of the SB
are constant during the time period between two successive glitch-events, then the equation describing the evolution of torque interaction between the SB and the ambient medium, i.e. Eq. (i/9), can be integrated straightforwardly.

\begin{figure}
\centering {\hspace*{0.75cm}
\includegraphics*[angle=-0, width=6.5cm]{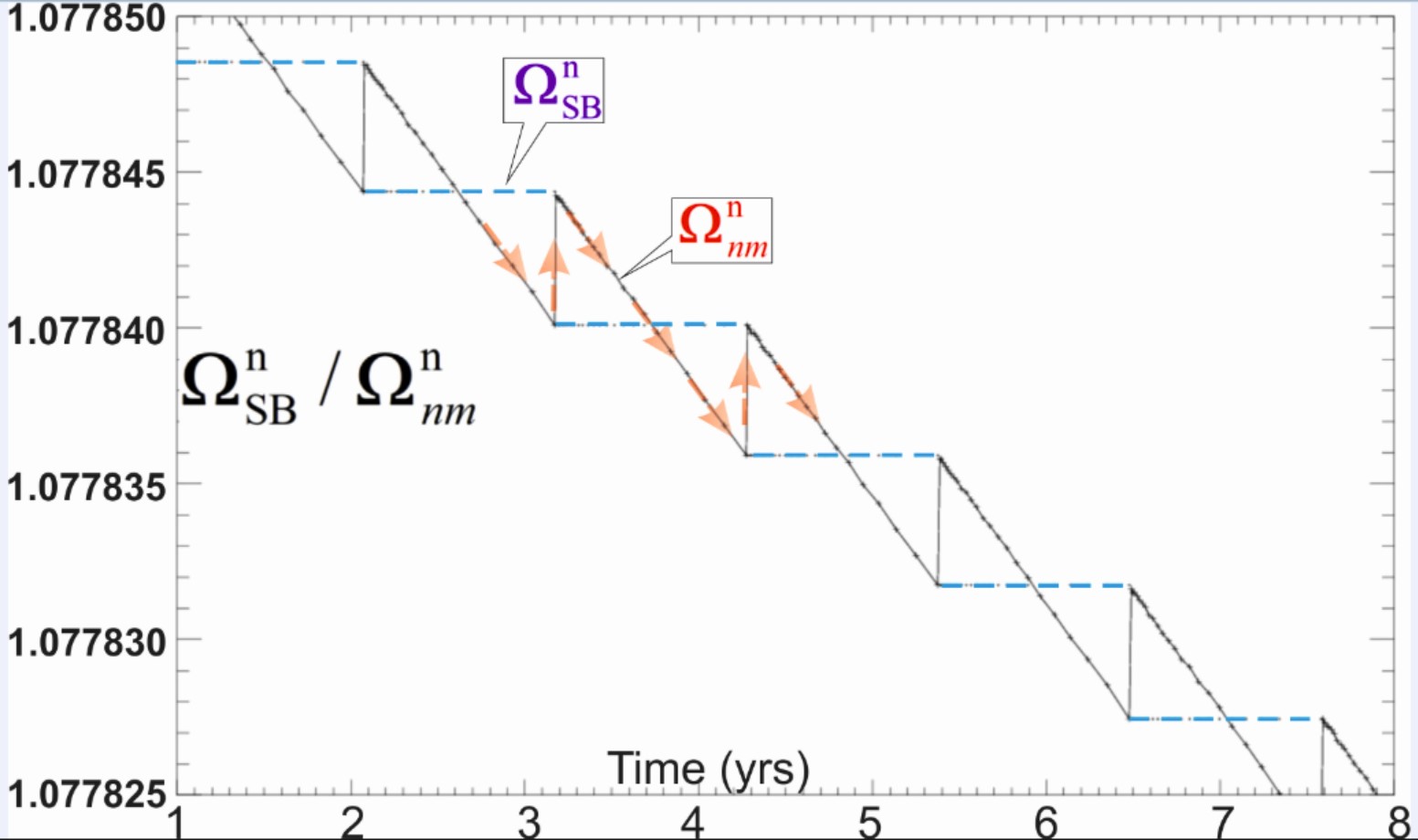}\\
\vspace*{0.25cm}
\includegraphics*[angle=-0, width=6.5cm]{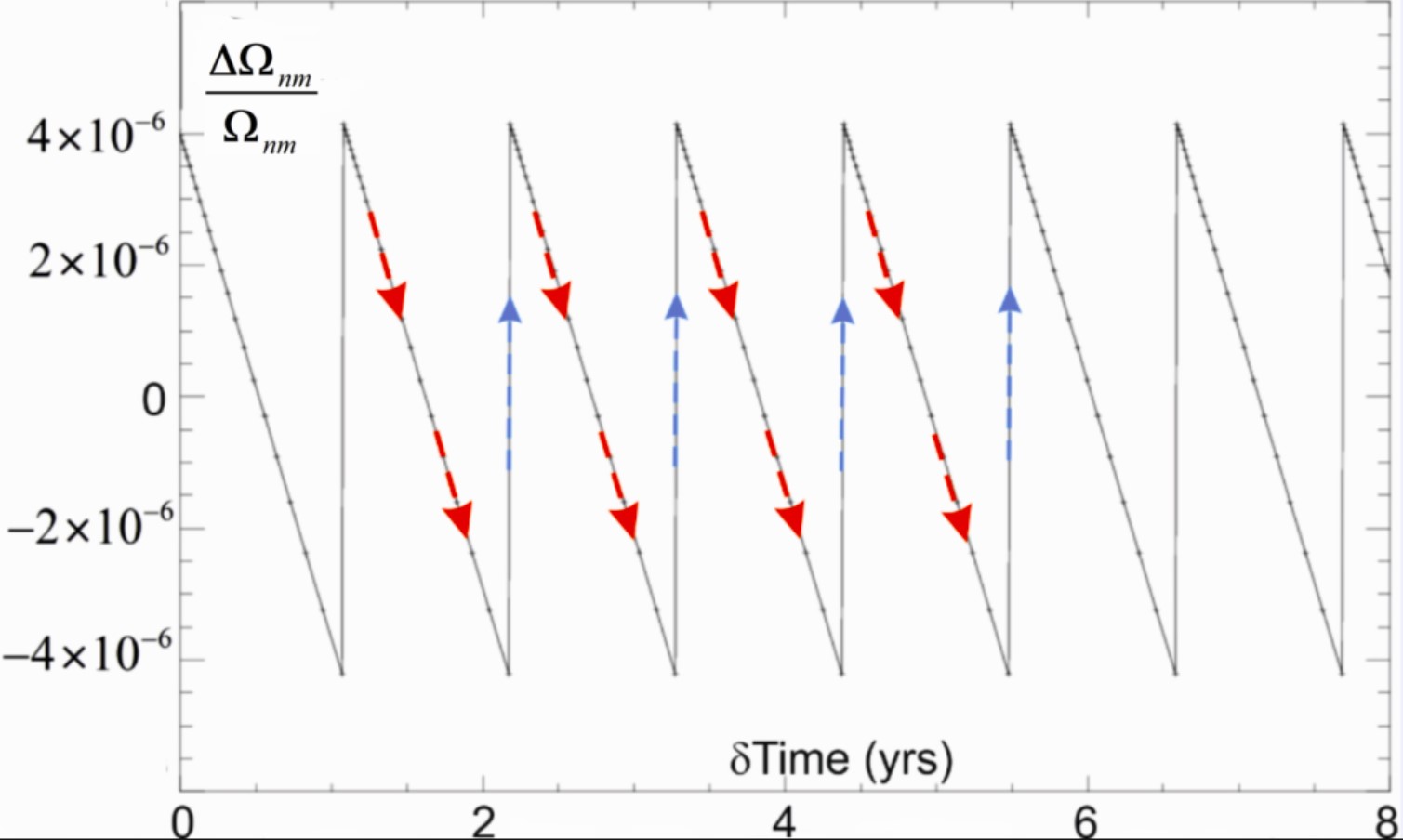}
}
\caption{\small The time-development of both the rotational frequencies of the SB ($\Omega^{now}_{SB}$) and that of the ambient medium
($\Omega^{now}_{}nm$) at the present epoch of the Vela pulsar are shown (upper panel).
Here time is in years and $\Omega$ in $70~/s$ units.  The term "now" resembles the present age
of the pulsar, i.e. roughly 11000 yrs.
Note the step function pattern of $\Omega^{now}_{SB},$ whereas the decrease of $\Omega^{now}_{nm}$ between two successive glitch-events evolve in a continuous manner.
In the lower panel the slow decay (red arrows) and the prompt increase (blue arrows) of
$\Omega^{now}_{nm}$ are shown.
Note that ${\hat{\Delta\Omega}}/\Omega = (\Omega^{max}- \Omega^{min})/\Omega$ falls within the
present observed values of the Vela pulsar.
Except the time-interval of glitch-events   the difference between $\Omega_{SB}$ and
$\Omega_{nm}$ is macroscopically insignificant on the years and their behavior would look  like identical for external observers.}
  \label{XXX}
\end{figure}

\begin{figure}
\centering {\hspace*{0.75cm}
\includegraphics*[angle=-0, width=6.5cm]{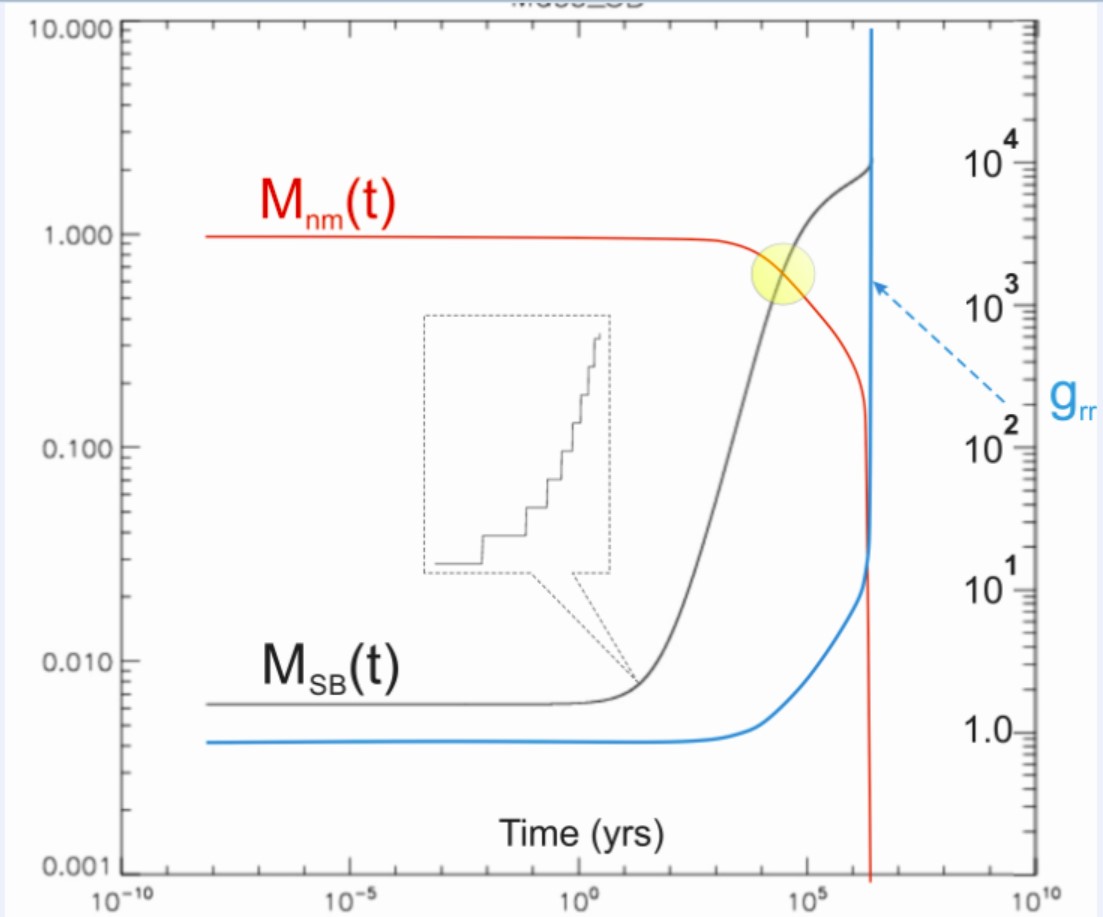}
}
\caption{\small The long-term  evolution of the  mass of the SB ($\calM_{SB}$/black line) and that of the ambient dissipative medium ($\calM_{nm}$/red line). As shown in the zoomed-window, the evolution of both masses proceed discretely with time rather than continuously. After roughly 200000 yrs the mass of embryonic SB surpasses the initial mass of pulsar and the object becomes increasingly red-shifted as indicated by the metric coefficient $g_{rr}$ (blue-colored line) to completely disappear from our observational windows after approximately several million years. The final mass of the SB is 2.8~$MSun$. The yellow circle represents the present state of the pulsar.
}  \label{XXX}
\end{figure}

\begin{figure}
\centering {\hspace*{0.75cm}
\includegraphics*[angle=-0, width=6.5cm]{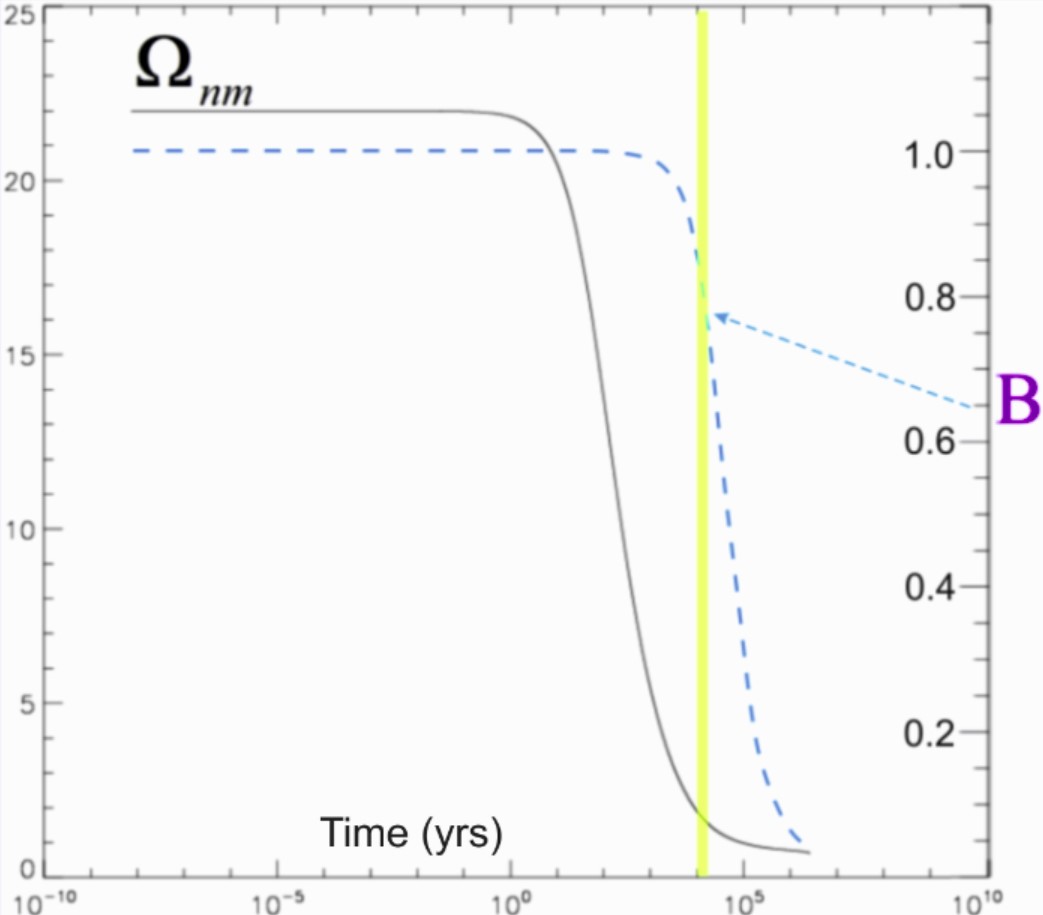}
}
\caption{\small The long-term  evolution of the  both the angular frequency of the
ambient medium in $\Omega_0=70/s$ units and of the magnetic field in $10^{13}$ Gauss. As mentioned in the text,  the magnetic field and inertia of the ambient medium in pulsar must
asymptotically decay to zero in order to fit present-day observation.
}  \label{XXX}
\end{figure}
\begin{figure}
\centering {\hspace*{0.0175cm}
\includegraphics*[angle=0, width=8.275cm]{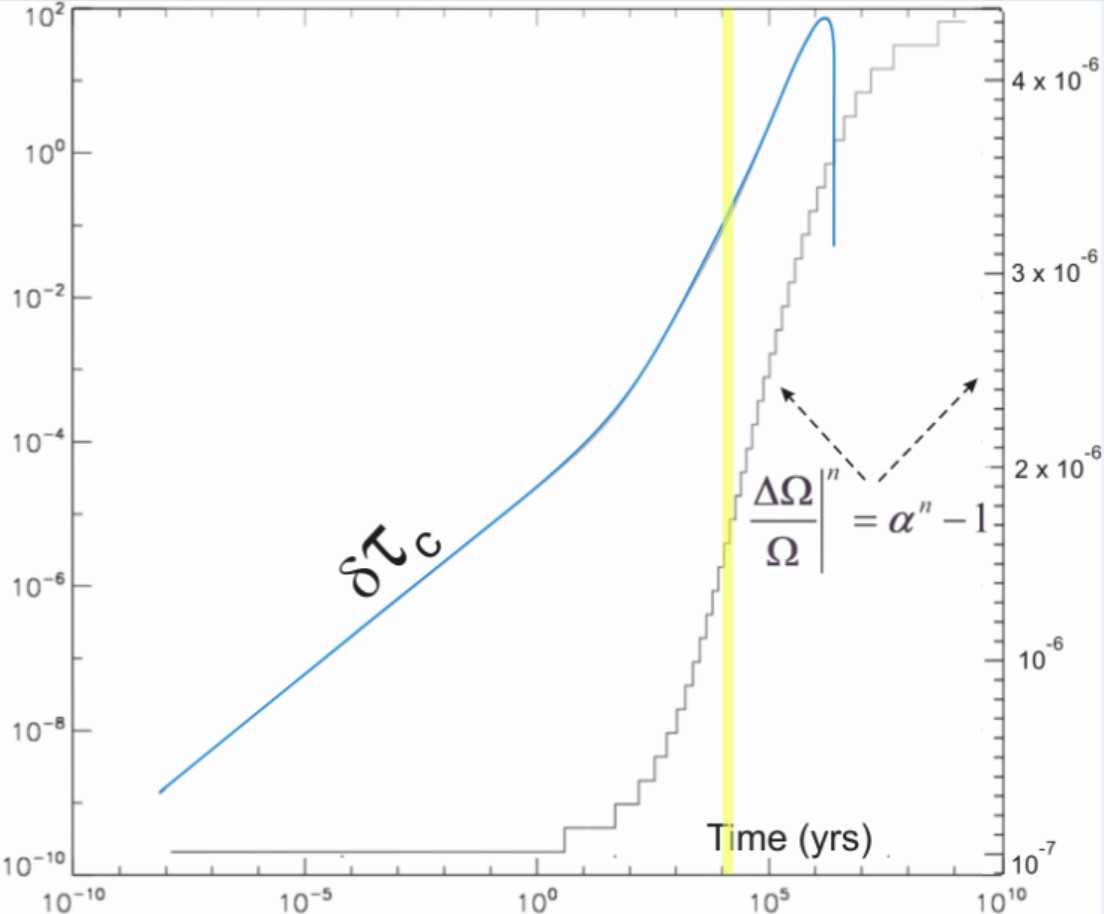}
}
\caption{\small The elements of the sequence $\alpha_c(t) = {\alpha^n_c}$ for
$n=0 \rightarrow N^\infty(\gg 1)$ and the corresponding
time interval between two successive glitch events, $\delta \tau_c,$ versus time in units of years are shown.
The remarkable change in the behavior of $\delta \tau_c$ for $t\geq 10^6$ yrs
is a consequence of the dramatic decrease of the inertia of the ambient dissipative medium
towards the end of the pulsar's lifetime.
}  \label{XXX}
\end{figure}
Based on the numerical solution of the equations listed in Eq. (10), I show in Fig. (5) the time-development of the rotational frequencies $\Omega^n_{SB}$  of the SB and of the ambient dissipative medium  $\Omega^n_{nm}.$ The top panel shows the prompt spin-down/spin-up of both fluid-components in units of 70 rotations per second in interval of approximately two years and 11000 years after the birth of the pulsar. These results agree excellently with  the revealed observational data of the Vela pulsar. In the lower panel the continuous spin-down and the prompt spin-up of the ambient medium versus time are shown, which again agrees well
with observations of the Vela pulsar.\\
The numerical solutions of the set of Equations (10) on the cosmological time-scale are shown in Figs. (6, 7) and (8). Here the long-term process of the metamorphosis is obvious: while
the total energy of the SB is growing, the mass of the ambient medium is diminishing to
finally disappear after roughly $5\times 10^6\,$yrs, or equivalently, after the object has
suffered of roughly forty  million glitches. At this time the object attains its maximum
compactness, i.e. $ \alpha_S \rightarrow 1$, it sinks deeply in spacetime  as the metric coefficient, $g_{rr},$ in the vicinity of the surface becomes extraordinary large.
This implies that the redshift would be very large and therefore the object
disappears from all possible observational windows.
In the absence of external forces, thermal,  magnetic and rotational energies of pulsars
must decrease during their lifetimes. Assuming newly born pulsars to initially rotate
with $\Omega_{nm}(t=0) = 1540 /s$ and that their initial magnetic fields are  $B(t=0) = 10^{13}\,$Gauss, then
the present calculations predict that these energies would be vanishingly small when
the object becomes several million years old (see Fig. 7).\\
Finally, given that the  initial mass and inertia of SBs are comparable to those of  nucleons
in atomic nuclei and therefore negligibly small compared to those of the ambient medium,
then the time interval between two succussive glitches must be respectively short,
though it becomes increasingly longer as the SBs age. Obviously, after one million years,
the decreasing-rate  of the inertia of the ambient medium
surpasses that of the magnetic field, thereby  shortening the time interval between successive glitches considerably (see Eq. 10 and Fig. 8).

\section{Summary and discussion}
In this article a new scenario for explaining the origin and dynamics of the glitch-phenomena observed in pulsars and neutron stars have been presented. Accordingly, birth of pulsars
most likely is associated with the formation of embryonic SBs, whose cores are made of supranuclear superconducting gluon-quark superfluid of uniform density $\rho = 3\times \rho_0.$ Following a well-defined  discrete quantum scheme in combination with a scalar field, the mass and inertia
of an SB should grow with time to finally metamorphose the entire pulsar into an invisible super baryon.
The SB-pulsar system is strikingly similar to a terrestrial rotating ${}^4$Helium inside a container. If both components are set to initially rotate with the same frequency and subsequently increasing/decreasing the frequency of the container in a continuous manner, then the enclosed helium-superfluid would increase/decrease its frequency discretely and promptly,
as dictated by the Onsager-Feynman equation.\\
The analogy to the SB-pulsars system is obvious: The SB corresponds to ${}^4$Helium,
whereas ambient~dissipative~medium to the solid container.\\
However, the situation in the SB-pulsar is still much more difficult as it requires exact determination of the permitted quantum energy levels of a gravitationally bounded super-baryon, which is completely unkown.\\
In the here-proposed scenario, the nuclear shell model is adopted to describe these energy levels, which are assumed to coincide with the angular frequency levels. A glitch event here would correspond to a transition from one energy level to the next, or equivalently, from one spin frequency to the next lower one. These transitions are provoked mainly by the two mechanisms:
\begin{enumerate}[leftmargin=*,labelindent=0pt,label=\bfseries \arabic*.]
  \item Los of rotational energy of the ambient medium through the magnetic torque interaction of  pulsar with their surrounding
  \item Enhancement of the effective mass-energy of the SB by the scalar field. Once the entire pulsar has metamorphosed into an SB, the mass of the object would double, provided the pulsar has the compactness parameter $\alpha_s=1/2$ initially.  Hence scalar fields act as compressors to the nuclear matter of the ambient media and enforce their transition into incompressible superconducting gluon-quark superfluids.
\end{enumerate}
  In fact, based on the discussion of Sec. (2), the solution procedure here can be used to estimate the relative width  of the BL as follows:
  \beq
     \DD{\delta R^n_{BL}}{R^n} = \sqrt{\DD{\Omega^n}{\Omega^{n+1}}} -1
     \approx \DD{\Delta \Omega}{\Omega}\approx
      \left\{
       \begin{array}{ll}
               10^{-8} & Crab \\
               10^{-6} & Vela,
             \end{array}
  \right.
  \eeq
  where $R^n$ is the radius of the SB at time $t^n.$ Obviously, as
  $\Omega^0/\Omega^{1}\approx 1 ,$ but much smaller than $ \Omega^{N}/\Omega^{N+1}$ for $N>>1$ and that
   $\delta R^{(n=0)}_{BL} \approx 1$ fm, then  $\delta R^n_{BL}$ is practically
   a measure for the length scale over which the residual of the strong nuclear
   force is communicated to the surrounding matter. As it is shown in Fig. (8), the magnification
   of such a quantum mechanical phenomenon becomes even crass when $t > 10^6$ yrs, as then  $\delta R^n_{BL}$ stretches out over several centimeters. While a thorough investigation of the strong nuclear force is beyond the scope of the present paper, relating
   $\delta R^n_{BL}$ to this fundamental force is inevitable consequence when coupling the
   observed glitch phenomena in pulsars to the evolution of embryonic SBs at their centers.
However, it remains unclear whether glitch-events in pulsars can be really organized in a "magic " mathematical sequence  that most pulsars evolve accordingly.  Assuming such a  sequence to exist, then we may use theoretical and observational arguments to constrain its properties (see List 7).
In fact, using the Onsager-Feynman equation, a lower-bound on $S^0$ can be calculated. Specifically, assume a pulsar to have been born with $\Omega_0=1400/s$ and
$B=10^{13}$ Gauss. Requiring that energy loss via the emission of magnetic dipole radiation to be smaller than that liberated rotational energy during a glitch-event, then we obtain $\Delta \Omega/\Omega \geq 3.8 \times 10^{-10}.$ In this case, the minimum natural number of vortices that could be expelled reads:
\beq
\delta N \geq \DD{2m}{h} S_{SB} (\Delta \Omega) = 5.32\times 10^{-4} S^0,
\eeq
 which is equivalent to require $ S^0 \geq 1.88 \times 10^3~cm^2.$\\
 Unlike classical models of glitches, the expelled vortices here are subject to  destructive frictional effects exerted by the surrounding dissipative medium and therefore they may diffuse out long before affecting the crust. It should be noted that even in the ideal case, vortex lines may interact with each other, reconnect and turn turbulent \citep[see][and the references therein]{Baggaley2014}. Therefore, it is unlikely that km-long vortex lines in SB-pulsar systems would behave differently.
 As a consequence, their
 spin-up effect during a glitch-event would be rather negligible, unless the dimensions of the SB becomes of a significant size.
 Assuming the total rotational energy to be stored in a network of vortices, then  the energy
 of one single vortex is upper-bounded by:
 \beq
       E^{kin}_V \leq (\DD{h}{4m} \DD{I \Omega}{S_\star})\approx 10^{31}~erg.
 \eeq
 Thus the minimum number of vortices needed to be expelled by the SB in order to be observationally noticeable would be $N \geq I\Omega \dot{\Omega}/E^{kin}_V \approx 10^7.$
 In the case of the Vela pulsar this implies that the radius of the enclosed SB must be larger than $10^{-2}R_\star,$ i.e. of a macroscopic length scale.\\
 In the Crab case, the radius of the SB compared to that in the Vela can be inferred  from the time-interval between two successive glitch-events as described in Eq. (11).
 Inserting the present values of $\Omega,~ \dot{\Omega}$ of both pulsars, then we obtain:
 \beq
   \delta \tau ^{Crab} \approx \DD{1}{2\beta} \DD{I}{B^2} (10^{-7}), \textrm{ and }
   \delta \tau ^{Vela} \approx \DD{1}{2\beta} \DD{I}{B^2} (10^{-5}).
   \eeq
  Recalling that the magnetic fields and the time-interval between two successive glitch-events in both system are of the same order, then we conclude that the inertia of the ambient dissipative medium in both pulsars may be related to each other as follows:
 \beq
 I^{Vela} \approx 10^{-2} I^{Crab}.
 \eeq
 This implies that the radius of the SB inside the Vela pulsar is at least 5/2 larger than
 the one inside the Crab, which is a reasonable outcome, when taking into account the difference in mass and age of both pulsars.\\
 It should be noted however that if the size of the embryonic SB is relatively small,
then the number of ejected vortices would be respectively small and therefore may diffuse out quickly without causing  an observationally significant spin-up of the crust.\\
 Finally, while much more efforts are needed still to refine and further constrain the universal "magic" sequence describing the evolution of  glitch-events in pulsars and neutron stars, I do think that the present scenario is promising as it is capable of comfortably accommodating observational data to provide answers related to the quantum origin and dynamics of glitches in pulsars.
 Specifically, glitch-observations appear to support the formation and evolution of embryonic
  SBs at the centers of pulsars, whose mass-growth follows a well-defined quantum sequence. The existence of such embryonic objects are conditioned to the occurrence of a phase transition form dissipative nuclear fluid into incompressible superconducting gluon-quark superfluid, which in turn requires a universal scalar field $\phi$ as a catalyst.  \\
    As quarks make at most $2\%$ of baryon mass,  we may simply view the $\phi-$field as a universal mechanism for enhancing the density of communication channels between quarks
    inside SBs, which in turn enhances its effective mass.

 \noindent Finally, similar to quarks in nucleon of atomic nuclei,  it was
 argues that stellar-size SBs cannot exist in free-space \citep{Witten1984}. However, the present scenario suggests that these objects must be the natural consequence of dying
 pulsars and neutron stars in an ever expanding universe. Such stellar-size SBs must
 be completely invisible due to their maximum compactness, which makes them excellent BH-candidates. Moreover, the role of the dark energy appear to be vital for metamorphosing pulsars into stellar-size super-baryons.